\documentclass{jpp}
\usepackage{ifpdf}

\usepackage{amssymb}
\usepackage{color}

 \ifpdf
  \usepackage{natbib}
  \fi

\def\beq{\begin{equation}}
\def\eeq{\end{equation}}
\def\een#1{\label{#1} \end{equation}}
\def\beqa{\begin{eqnarray}}
\def\eeqa{\end{eqnarray}}
\def\ean#1{\label{#1} \end{eqnarray}}
\def\pd#1#2{\frac{\partial{#1}}{\partial{#2}}}
\def\od#1#2{\frac{d{#1}}{d{#2}}}
\def\eqref#1{(\ref{#1})}
\def\nnn{\nonumber \\}
\def\eps{\varepsilon}
\def\etal{\textit{et al.\/}}
\def\sech{{\rm sech}}

\begin{document}

\title[Modified KdV solitons in plasmas at supercritical densities]%
    {Modified Korteweg-de Vries solitons at supercritical densities in
    two-electron temperature plasmas}
\author[F. Verheest, C. P. Olivier \& W. A. Hereman]
{F\ls r\ls a\ls n\ls k\ns V\ls e\ls r\ls h\ls e\ls e\ls s\ls t $^{1,2}$
    \footnote{Email address for correspondence: frank.verheest@ugent.be},
C\ls a\ls r\ls e\ls l\ns P.\ns O\ls l\ls i\ls v\ls i\ls e\ls r $^3$ and
W\ls i\ls l\ls l\ls y\ns A.\ns H\ls e\ls r\ls e\ls m\ls a\ls n $^4$ }
\pubyear{2016}
\affiliation{$^1$Sterrenkundig Observatorium, Universiteit Gent,
    Krijgslaan 281, B--9000 Gent, Belgium \\[\affilskip]
$^2$School of Chemistry and Physics, University of KwaZulu-Natal,
        Durban 4000, South Africa \\[\affilskip]
$^3$Space Science, South African National Space Agency, P.\ O.\ Box 32,
    Hermanus 7200, South Africa \\[\affilskip]
$^4$ Department of Applied Mathematics and Statistics, Colorado School of Mines,
    Golden, CO 80401--1887, USA }



\maketitle

\begin{abstract}
The supercritical composition of a plasma model with cold positive ions in the
presence of a two-temperature electron population is investigated, initially by
a reductive perturbation approach, under the combined requirements that there
be neither quadratic nor cubic nonlinearities in the evolution equation.
This leads to a unique choice for the set of compositional parameters and a
modified Korteweg-de Vries equation (mKdV) with a quartic nonlinear term.
The conclusions about its one-soliton solution and integrability will also be
valid for more complicated plasma compositions.
Only three polynomial conservation laws can be obtained.
The mKdV equation with quartic nonlinearity is not completely integrable, thus
precluding the existence of multi-soliton solutions.
Next, the full Sagdeev pseudopotential method has been applied and this allows
for a detailed comparison with the reductive perturbation results.
This comparison shows that the mKdV solitons have slightly larger amplitudes
and widths than those obtained from the more complete Sagdeev solution and that
only slightly superacoustic mKdV solitons have acceptable amplitudes and
widths, in the light of the full solutions.
\end{abstract}

\section{Introduction}

Nonlinear solitary waves in various plasma models and compositions have been
investigated for the last half century, both theoretically and observationally.
Theoretical descriptions were initially based on reductive perturbation
techniques [\cite{Zabusky1965,WasTan}] and, almost contemporarily, on the
Sagdeev pseudopotential method [\cite{Sagdeev}].
This has resulted in a vast body of literature which is hard to cite in a way
which would do it justice.
We therefore only refer to those papers needed for the understanding or
illustration of our present endeavour.

Reductive perturbation methods have the advantages of both flexibility and
algorithmic procedures in exploring many different models.
Yet, they are restricted to weakly nonlinear waves by the iterative way of
working through the (asymptotic) expansions which are only valid for
sufficiently small amplitudes.
This precise aspect is difficult to quantify and its limitations are often
disregarded in numerical illustrations.
For acoustic solitary modes, the archetype is the Korteweg-de Vries (KdV)
equation [\cite{KdV}], initially established for surface waves on shallow
water, but much later found to have applications in many other fields of
physics [\cite{Zabusky1965,Miura}], particularly in plasma physics
[\cite{WasTan}].

On the other hand, the Sagdeev pseudopotential procedure [\cite{Sagdeev}] is
more difficult to work through because it requires at the intermediate stages
integrations and inversions to express all dependent variables in terms of a
single one.
Finding the latter is not always obvious, let alone possible.
Its advantage, however, is that it is not limited to solitary waves of small
amplitudes, but admits large though bounded solutions, given the various
restrictions imposed by the model.
Some models that admit solitary waves can be treated by both methods.
In those cases, the expansion of the Sagdeev pseudopotential to its lowest
significant orders is instructive because it offers an insight in the
acceptability of the reductive perturbation results by determining the
deviation from the fully nonlinear solutions.
This will be illustrated for the model treated in this paper.

The relative success of reductive perturbation theory in describing nonlinear
wave problems is based on a separation of fast and slow timescales and of
linear and nonlinear effects.
Ideally, this leads to a balance between nonlinearity and dispersion enabling
the emergence of stable solitary waves that propagate unchanged in time and
space.
These waves are characterized by nonlinear relations between amplitude, width,
and propagation speed [\cite{Drazin}].
In addition, KdV solitons have remarkable interaction properties.
Indeed, if slower solitons are overtaken by faster ones they both emerge from
the collision unaltered, apart from a phase shift [\cite{Zabusky1965}].
The application of reductive perturbation theory requires two key elements: a
proper stretching to rearrange the independent variables (essentially a
co-moving coordinate at the linear phase speed and a slow time scale), plus a
suitable expansion of the dependent variables.
The linear dispersion properties govern the choice of stretching
[\cite{Davidson}], which, in turn, determines the form of the evolution
equation one obtains, when coupled to the expansion scheme.

For simple wave problems, like the nonlinear description of an ion-acoustic
soliton in an electron-proton plasma, most of the compositional parameters are
fixed or eliminated by a proper normalization, and the result is the ubiquitous
KdV equation [\cite{WasTan}], with a quadratic nonlinearity, reflecting the
ordering between the scaling of the independent variables and the parameter
governing the expansion of the dependent variables.
When the plasma model becomes more involved, there are so-called critical
choices for the compositional parameters which annul the coefficient of the
nonlinear term in the KdV equation leading to a undesirable linear equation.
In other words, the combination of stretching and expansion used must then be
adapted to account for nonlinear effects of higher degree.
This is easiest done for the stretching and leads to the modified KdV (mKdV)
equation [\cite{WatTan,Buti,Watanabe}] with a cubic nonlinearity.

An interesting question which then arises is whether one can take this
procedure to a higher level.
Indeed, one might wonder if for complicated enough plasma models the
coefficients of both the quadratic \textit{and} the cubic nonlinearities can
be annulled simultaneously for a specific and clearly restricted set of
compositional parameters.
Obviously, for many models and soliton types such supercritical compositions
will be impossible.
Some aspects of this issue have been discussed before
[\cite{TagareCrit,TagareKdV}] in an effort to establish classes of wave
problems for which supercriticality cannot occur.

However, there are many situations where supercriticality is possible, even
though the model might become rather constrained and is consequently not
easily physically realisable.
Nevertheless, there are several aspects of this problem which merit closer
attention because they lead to a type of KdV equation which has not been much
derived in the plasma physics literature, although it was on the radar of the
discoverers of solitons [\cite{Kruskal-etal-1970,Zabusky1967,Zabusky1973}] and
has been quoted in more mathematically inclined studies as one of the
higher-degree extensions of the KdV family of equations
[\cite{Drazin,Wazwaz2005,Wazwaz2008}].

In the present paper, we investigate a rather simple plasma model with cold
positive ions in the presence of a two-temperature electron population, and
show that it can indeed exhibit supercritical behavior.
Although there is no compositional freedom left for the model under
investigation, the conclusions are instructive and will remain valid for
classes of more complicated plasmas, with, e.g., four rather than three
species, yet at the cost of more complicated algebra [\cite{Carel}].
Moreover, this three-constituent plasma model has also been studied via the
Sagdeev pseudopotential method [\cite{DoubleBoltz}], though for generic values
of the composition, with the focus on changes in electrostatic polarity of the
resulting modes, and related issues.
We will thus be able to compare the reductive perturbation and Sagdeev
pseudopotential treatments, and infer some of the limitations of the former, in
terms of its numerical validity.

The paper is structured as follows.
The reductive perturbation analysis is presented in Section 2, showing that we
can indeed have supercritical densities and temperatures, leading to an mKdV
equation with a quartic nonlinearity.
Its soliton properties are then investigated in Section 3.
In Section 4 the problem is treated with the Sagdeev pseudopotential approach,
allowing for a comparison with the reductive perturbation results in the weakly
nonlinear case.
Finally, conclusions are summarized in Section 5.

\section{Reductive perturbation formalism at supercritical densities}

\subsection{Model equations}

We consider a three-component plasma comprising cold fluid ions and two
Boltzmann electron species at different temperatures
[\cite{NishTaj,DoubleBoltz}].
The basic equations are well known, and consist of the continuity and
momentum equations for the cold ions, and Poisson's equation coupling the
electrostatic potential $\varphi$ to the plasma densities.
Restricted to one-dimensional propagation in space and written in normalized
variables, the model reads
\begin{eqnarray}
&& \pd{n}{t} + \pd{}{x} \left( n u \right) = 0, \label{cont} \\
&& \pd{u}{t} + u\, \pd{u}{x} + \pd{\varphi}{x} = 0, \label{mot} \\
&& \pd{^2 \varphi}{x^2} + n - f \exp[\alpha_c \varphi]
    - (1-f) \exp[\alpha_h \varphi] = 0. \label{pois}
\end{eqnarray}
Here $n$ and $u$ refer to the ion density and fluid velocity, respectively, and
$f$ is the fractional charge density of the cool electrons.
The temperatures $T_c$ and $T_h$ of the Boltzmann electrons are expressed
through $\alpha_c = T_{\rm eff}/T_c$ and $\alpha_h = T_{\rm eff}/T_h$ for the
cool and hot species, respectively, whereas the effective temperature is given
by $T_{\rm eff} = T_c T_h/[f T_h + (1-f)T_c]$, such that
$f \alpha_c + (1-f) \alpha_h = 1$.
In this description densities are normalized by their undisturbed values (for
$\varphi=0$), velocities by the ion-acoustic speed in the plasma model,
$c_{ia}=\sqrt{\kappa T_{\rm eff}/m_i}$, the electrostatic potential by
$\kappa T_{\rm eff}/e$, length by an effective Debye length,
$\lambda_D = \sqrt{\eps_0 \kappa T_{\rm eff}/(n_{i0} e^2)}$, and time by the
inverse ion plasma frequency,
$\omega_{pi}^{-1} = [n_{i0} e^2/(\eps_0 m_i)]^{-1/2}$.
Hence, the dependent and independent variables as well as the parameters in
\eqref{cont}--\eqref{pois} are dimensionless.

Various KdV-like equations have been studied in a great variety of plasma
models.
We briefly review the two equations that are most relevant to this paper but
are widely studied in the relevant plasma physics literature
[\cite{VerheestASSL}] and elsewhere [\cite{Drazin}].
The standard KdV equation is of the form
\begin{equation}\label{kdv}
\pd{\psi}{\tau} + B\,\psi\,\pd{\psi}{\xi} + \pd{^3\psi}{\xi^3} = 0,
\end{equation}
where $\xi$ and $\tau$ refer to the stretched space and time variable,
respectively, to be defined later, $\psi$ is the relevant lowest-order term in
an expansion of $\varphi$ and the coefficients of the slow time variation term
($\partial\psi/\partial\tau$) and the dispersive term
($\partial^3\psi/\partial\xi^3$) have been rescaled to unity.
This can be done without loss of generality for these coefficients were
strictly positive.
The coefficient $B$ of the quadratic nonlinearity has in principle no fixed
sign as it depends on the details of the plasma model.
In the generic case $B\not= 0$, but when the plasma composition is critical,
$B=0$ and the analysis has to be adapted accordingly.
Doing so, yields in principle the well-studied mKdV equation [\cite{Wadati}],
\begin{equation}\label{mkdv}
\pd{\psi}{\tau} + C\,\psi^2\,\pd{\psi}{\xi} + \pd{^3\psi}{\xi^3} = 0,
\end{equation}
with a cubic nonlinearity.
As has been shown for certain modes and plasma compositions
[\cite{TagareCrit,TagareKdV}], under the conditions that the dispersion law is
adhered to and $B=0$, it is not easy to make $C=0$ for it implies severe
restrictions on the compositional parameters.

However, as will be seen, the model with two Boltzmann electrons and cold ions
allows one to have both $B=0$ and $C=0$ at the cost of the compositional
parameters $f$, $\alpha_c$ and $\alpha_h$ being completely fixed.
We will call this a supercritical composition [\cite{TagareKdV}], which might
not easily be realized in practice, yet gives an insight in the special
properties of such a model.
In particular, its reductive perturbation analysis leads to a modified KdV
equation with a \textit{quartic} nonlinearity,
\begin{equation}\label{skdv}
\pd{\psi}{\tau} + D\,\psi^3\,\pd{\psi}{\xi} + \pd{^3\psi}{\xi^3} = 0,
\end{equation}
for which integrability issues and solitary wave solutions (solitons) will be
discussed below.
Although \eqref{skdv} appears to be new in connection with equations
\eqref{cont}--\eqref{pois}, it has received some attention in the early
development of soliton theory
[\cite{Kruskal-etal-1970,Zabusky1967,Zabusky1973}] and, on various occasions,
has resurfaced in the mathematical physics literature
[\cite{Wazwaz2005,Wazwaz2008}].

\subsection{Reductive perturbation analysis at supercritical densities}

There is no point in first deriving the usual KdV equation \eqref{kdv}, thus
finding $B$ explicitly, then assuming that $B=0$, changing the stretching and
deriving \eqref{mkdv}, with the expression for $C$, so that $C=0$ (still under
the restriction that $B=0$) requiring yet another stretching.
These steps are well known, and there is a plethora of papers and books where
this procedure is illustrated [\cite{VerheestASSL}].

Instead, we start from a stretching which will generate \eqref{skdv} right
away.
The properties of the stretching can be determined in several ways.
Here we use the scaling properties of \eqref{skdv}.
A comparison of the first and last terms indicates that
$\partial/\partial\tau \sim \partial^3/\partial\xi^3$, and from the middle and
the last terms one gets $\psi^3 \sim \partial^2/\partial\xi^2$.
Using a standard expansion, we take
\begin{eqnarray}\label{expan}
n &=& 1 + \eps n_1 + \eps^2 n_2 + \eps^3 n_3 + \eps^4 n_4 + ... , \nnn
u &=& \eps u_1 + \eps^2 u_2 + \eps^3 u_3 + \eps^4 u_4 + ... , \nnn
\varphi &=& \eps \varphi_1 + \eps^2 \varphi_2  + \eps^3 \varphi_3
    + \eps^4 \varphi_4 + ... \, .
\end{eqnarray}
We strive to obtain a nonlinear evolution equation in $\psi=\varphi_1$ and thus
$\partial/\partial\xi \sim \eps^{3/2}$ and
$\partial/\partial\tau \sim \eps^{9/2}$, leading to the stretched variables
\begin{equation}\label{stret}
\xi = \eps^{3/2}(x-t), \qquad \tau = \eps^{9/2} t.
\end{equation}
This means that \eqref{cont} and \eqref{mot} will yield terms to order
$\eps^{5/2}$, $\eps^{7/2}$ and $\eps^{9/2}$ which can be integrated with
respect to $\xi$, the derivatives with respect to $\tau$ only appearing at the
order $\eps^{11/2}$.
In these integrations it is assumed that for solitary waves all dependent
variables and their partial derivatives with respect to $\xi$ vanish for
$|\xi|\to \infty$.
Thus, the intermediate results are
\begin{eqnarray}\label{econt}
n_1 &=& u_1, \nnn
n_2 &=& u_2 + n_1 u_1, \nnn
n_3 &=& u_3 + n_1 u_2 + n_2 u_1,
\end{eqnarray}
and
\begin{eqnarray}\label{emot}
u_1 &=& \varphi_1, \nnn
u_2 &=& \varphi_2 + \textstyle{\frac{1}{2}} u_1^2
    = \varphi_2 + \textstyle{\frac{1}{2}} \varphi_1^2, \nnn
u_3 &=& \varphi_3 + u_1 u_2
    = \varphi_3 + \varphi_1 \varphi_2 + \textstyle{\frac{1}{2}} \varphi_1^3.
\end{eqnarray}
Eliminating $u_1$, $u_2$ and $u_3$ from \eqref{econt} and \eqref{emot}
yields
\begin{eqnarray}\label{eres}
n_1 &=& \varphi_1, \nnn
n_2 &=& \varphi_2 + \textstyle{\frac{3}{2}} \varphi_1^2, \nnn
n_3 &=& \varphi_3 + 3 \varphi_1 \varphi_2 + \textstyle{\frac{5}{2}} \varphi_1^3.
\end{eqnarray}

This has to be linked to results from \eqref{pois} to order $\eps$, $\eps^2$
and $\eps^3$, before the Laplacian contributes to order $\eps^4$.
For notational brevity we introduce
\begin{equation}\label{Aa}
A_\ell = f\alpha_c^\ell + (1-f)\alpha_h^\ell
    = f(\alpha_c^\ell - \alpha_h^\ell) + \alpha_h^\ell \qquad
(\ell = 1, 2, 3, ...),
\end{equation}
so that \eqref{pois} leads to
\begin{eqnarray}\label{epois}
n_1 &=& A_1 \varphi_1, \nnn
n_2 &=& A_1 \varphi_2 + \textstyle{\frac{1}{2}} A_2 \varphi_1^2, \nnn
n_3 &=& A_1 \varphi_3 + A_2 \varphi_1 \varphi_2
    + \textstyle{\frac{1}{6}} A_3 \varphi_1^3.
\end{eqnarray}
Equating the expressions for $n_1$, $n_2$ and $n_3$ in \eqref{eres} and
\eqref{epois} leads to significant intermediate results:
\begin{eqnarray}\label{edisp}
&& \left(1-A_1\right)\varphi_1 = 0, \nnn
&& \left(1-A_1\right)\varphi_2 + \textstyle{\frac{1}{2}}
    \left(3-A_2\right)\varphi_1^2 = 0, \nnn
&& \left(1-A_1\right)\varphi_3 + \left(3-A_2\right)\varphi_1 \varphi_2
    + \textstyle{\frac{1}{6}} \left(15-A_3\right)\varphi_1^3 = 0.
\end{eqnarray}
In order to continue with $\varphi_1\not= 0$, the coefficients of the powers of
$\varphi_1$ in these equations need to vanish.
The first one, $A_1=1$, is nothing but the dispersion law, given the judicious
choices of $T_{\rm eff}$ and $c_{ia}$ in the normalization, and also in the
stretching \eqref{stret}.
The second one, $A_2=3$, is equivalent to the annulment of $B$ in the KdV
equation \eqref{kdv}.
The third one, $A_3=15$, is nothing but the annulment of $C$ in the mKdV
equation \eqref{mkdv}.

Before proceeding, one should be assured that these relations can be fulfilled
for $f$, $\alpha_c$ and $\alpha_h$.
Using \eqref{Aa} and slightly rewriting the conditions gives
\begin{eqnarray}\label{condit}
f(\alpha_c-\alpha_h) &=& 1 - \alpha_h, \nnn
f(\alpha_c^2-\alpha_h^2) &=& 3 - \alpha_h^2, \nnn
f(\alpha_c^3-\alpha_h^3) &=& 15 - \alpha_h^3,
\end{eqnarray}
from which it follows that
\begin{equation}\label{numval}
f=\textstyle{\frac{1}{6}}(3-\sqrt{6}), \qquad \alpha_c=3+\sqrt{6}, \qquad
    \alpha_h=3-\sqrt{6}.
\end{equation}
Eliminating $u_4$ between \eqref{cont} and \eqref{mot} at order $\eps^{11/2}$
and expressing all terms as functions of $\varphi_i$ yields
\begin{equation}\label{en4}
\pd{n_4}{\xi} = 2\pd{\varphi_1}{\tau} + \pd{\varphi_4}{\xi}
    + 3 \pd{}{\xi}(\varphi_1\varphi_3) + 3 \varphi_2 \pd{\varphi_2}{\xi}
    + \frac{15}{2}\, \pd{}{\xi}(\varphi_1^2\varphi_2)
    + \frac{35}{2}\, \varphi_1^3 \pd{\varphi_1}{\xi}.
\end{equation}
On the other hand, using the specific values \eqref{numval} rendering
effectively $B=0$ and $C=0$, from \eqref{pois} one finds to order $\eps^4$
that
\begin{equation}\label{pois4}
\pd{^2\varphi_1}{\xi^2} + n_4 - \varphi_4 - 3 \varphi_1\varphi_3
    - \frac{3}{2} \varphi_2^2 - \frac{15}{2} \varphi_1^2\varphi_2
    - \frac{27}{8} \varphi_1^4 = 0.
\end{equation}
Taking the derivative of this equation with respect to $\xi$ and eliminating
terms in $n_4$ and $\varphi_4$ yields the supercritical mKdV equation with a
quartic nonlinearity:
\begin{equation}\label{eskdv}
\pd{\varphi_1}{\tau} + 2 \varphi_1^3 \pd{\varphi_1}{\xi}
    + \frac{1}{2}\, \pd{^3\varphi_1}{\xi^3} = 0.
\end{equation}
Related results have been obtained by \cite{DasSen} for the same plasma model
but including relativistic effects and three-dimensional motion of the plasma
species.
This leads to variations of the Kadomtsev-Petviashvili (KP) equation
[\cite{Kadomtsev}], which reduce to the corresponding KdV equations when the
extra space-dimensional features are omitted.
A weakness of their approach is that they determine expressions equivalent to
our coefficients $B$, $C$ and $D$ of the nonlinear terms, and use the
conditions $B=C=0$ to derive supercritical KP and KdV equations, without
explicitly checking that this can indeed be done.
This oversight causes \cite{DasSen} to discuss the supercritical evolution
equations (of KP and KdV types) as if the coefficient of the quartic
nonlinearity (equivalent to our $D$) were a freely adjustable parameter of
either sign.
However, the values they should have computed from annulling the coefficients
of the quadratic and cubic nonlinearities are equivalent to \eqref{numval},
nonessential differences being due to a slightly different normalization.
In any case, $D=2$ follows, as expected, a positive parameter which cannot be
varied!

Before concluding this section, we stress that for more involved plasma
configurations allowing for $B=C=0$ (together with the usual conditions about
charge neutrality in the undisturbed configuration and the appropriate
dispersion law), an evolution equation with a quartic nonlinearity like
\eqref{skdv} and \eqref{eskdv} will be obtained, but with different
coefficients.
This implies that \textit{mutatis mutandis} the discussion and conclusions
about soliton solutions and integrability, obtained in the next section, will
still hold.

\section{Soliton solutions and integrability}

\begin{figure}
\centering
\includegraphics[width=120mm]{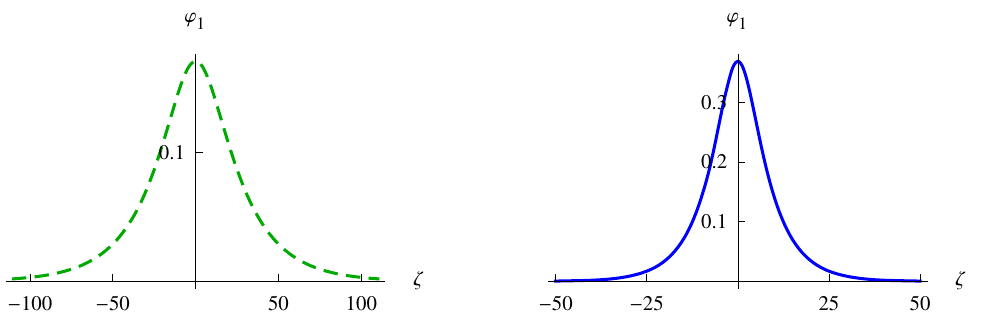}
\caption{Plot of KdV-type solitons for $W=0.001$ (left) and $W=0.01$ (right),
with amplitudes 0.171 and 0.368, respectively.}
\label{kdvSol}
\end{figure}

The discussion of \eqref{eskdv} involves two aspects: its solitary wave
(soliton) solutions, and its integrability, in particular, the lack of
so-called complete integrability.
A one-soliton solution is easily found by changing to a slightly superacoustic
coordinate,
\begin{equation}\label{zeta}
\zeta=\xi-W\tau,
\end{equation}
and using the tanh method [\cite{Malfliet,MalfHer}] or sech method
[\cite{Baldwin-etal-2004}].
Alternatively, as shown for the KdV equation in [\cite{Hereman2009}], one can
integrate \eqref{eskdv} twice which readily yields the solution
\begin{equation}\label{onesol}
\varphi_1 = \sqrt[3]{5 W} \sech^{2/3}\left(3\sqrt{\frac{W}{2}}\, \zeta\right).
\end{equation}
This solution is plotted in Fig.\ \ref{kdvSol} for $W=0.001$ (left) and
$W=0.01$ (right).
In the second case, the amplitude is already over the limit of what might be
acceptable in a reductive perturbation method resting on an expansion and
iterative procedure.
In an inertial frame this gives a soliton velocity of 1.01.

When we rewrite \eqref{eskdv} as
\begin{equation}\label{cons1}
\pd{\varphi_1}{\tau} + \frac{1}{2}\, \pd{}{\xi}
    \left( \varphi_1^4 + \pd{^2\varphi_1}{\xi^2} \right) = 0,
\end{equation}
the equation is of the form
\begin{equation}\label{conser}
\pd{\rho}{\tau} + \pd{J}{\xi} = 0.
\end{equation}
This is called a conservation law, with density $\rho$ and flux $J$, both
being functions of $\varphi_1$ and its derivatives with respect to $\xi$.
Because $\varphi_1$ and its derivatives go to zero as $|\xi|\to \infty$,
upon integration over the whole real line, one gets
\begin{equation}\label{consdens}
\int^{+\infty}_{-\infty} \pd{\rho}{\tau}\, d\xi
    + \int^{+\infty}_{-\infty} \pd{J}{\xi}\, d\xi
= \int^{+\infty}_{-\infty} \pd{\rho}{\tau}\, d\xi
    + J \, \big|_{-\infty}^{+\infty}
= \pd{}{\tau} \int^{+\infty}_{-\infty} \rho\, d\xi = 0.
\end{equation}
Consequently, $\int^{+\infty}_{-\infty} \rho\, d\xi$ remains constant when the
system evolves in time, and therefore $\rho$ represents the density of a
conserved integral.

Thus, \eqref{cons1} expresses that $\varphi_1$ is a conserved density.
As can straightforwardly be checked, two other independent conserved densities
and corresponding fluxes can be established with the method described in
[\cite{Schamel}],
\begin{eqnarray}
&& \pd{\varphi_1^2}{\tau} + \pd{}{\xi} \left[ \frac{4}{5}\, \varphi_1^5
    + \varphi_1\, \pd{^2\varphi_1}{\xi^2}
    - \frac{1}{2} \left( \pd{\varphi_1}{\xi} \right)^2 \right] = 0,
        \label{cons2} \\
&& \pd{}{\tau} \left[ \varphi_1^5
    - \frac{5}{2} \left( \pd{\varphi_1}{\xi} \right)^2 \right]
+ \pd{}{\xi} \left[ \frac{5}{4}\, \varphi_1^8
    - 10 \varphi_1^3 \left( \pd{\varphi_1}{\xi} \right)^2
    + \frac{5}{2}\, \varphi_1^4\, \pd{^2\varphi_1}{\xi^2} \right. \nnn
&& \qquad\qquad\quad
    +\, \left. \frac{5}{4}\, \left( \pd{^2\varphi_1}{\xi^2} \right)^2
    - \frac{5}{2}\, \pd{\varphi_1}{\xi}\, \pd{^3\varphi_1}{\xi^3} \right] = 0
        \label{cons3}.
\end{eqnarray}
One might think of these conservation laws as expressing conservation of mass,
momentum, and energy.
As an aside, we note that the building blocks of any conserved density or flux
belong together under the scaling properties of \eqref{eskdv}.
Noting that
\begin{equation}
\label{derivative}
\pd{}{\xi} \left(\varphi_1 \pd{\varphi_1}{\xi}\right)
= \varphi_1\, \pd{^2\varphi_1}{\xi^2} + \left( \pd{\varphi_1}{\xi} \right)^2,
\end{equation}
it is seen that, e.g., the building blocks of the flux in \eqref{cons2} are
those of the conserved density in \eqref{cons3}, barring the term
$(\partial/\partial\xi) (\varphi_1 \partial\varphi_1/\partial\xi)$ which can be
moved into the flux of \eqref{cons3}.
Full details about the construction of densities and the computation of fluxes
can be found in [\cite{Hereman-etal-2009,PooleHereman2011}].

Before continuing, it has been shown, historically first in a rather haphazard
way [\cite{Zabusky1965}], later more systematically [\cite{Miura}], that for
completely integrable equations like \eqref{kdv} (KdV) and \eqref{mkdv} (mKdV)
one can generate an infinite number of polynomial conserved densities.
This serves as one of the possible definitions of what is understood by
completely integrable nonlinear evolution equations.
Indeed, the existence of an infinite number of conserved densities is an
indicator that the evolution equation has a rich mathematical structure
resulting in the extraordinary stability of solitary waves and the elastic
collision property of ``solitons", a particle-like name appropriately coined by
Zabusky [\cite{Zabusky1965}].
Completely integrable nonlinear PDEs have remarkable features, such as a Lax
pair, a Hirota bilinear form, B\"acklund transformations, and the Painlev\'e
property.
They can be written as infinite-dimensional bi-Hamiltonian systems and have an
infinite number of conserved quantities, infinitely many higher-order
symmetries, and an infinite number of soliton solutions.

Modified KdV equations with a third-order dispersion term but nonlinearities of
degree higher than three, as in \eqref{skdv} or \eqref{eskdv}, are known to
have no more than three polynomial conservation laws
[\cite{Zabusky1967,Kruskal-etal-1970}], and none of those contain $t$ and $x$
explicitly.
Thus, there is a fundamental difference with the classical KdV and mKdV
equations which both have infinitely many independent polynomial conserved
densities and have long been known to be completely integrable.

With reference to \eqref{onesol}, we are in principle not allowed to use the
word ``soliton" since that name should be reserved for waves that collide
elastically.
Yet, adhering to common practice, we will continue to use soliton as a
shorthand for solitary wave.
In the absence of complete integrability, $N$-soliton solutions do not exist,
not even a genuine 2-soliton solution where a faster and taller soliton is seen
to overtake a slower and smaller one without distorting their shapes.
Further properties of the quartic KdV-type equation have been investigated by
\cite{Zabusky1973} and \cite{MM}.
In the latter paper, the authors discuss soliton stability and 2-soliton
interactions in an asymptotic sense for solitons of either widely different or
nearly equal amplitudes.

\section{Comparison with Sagdeev pseudopotential treatment}

The Sagdeev pseudopotential [\cite{Sagdeev}] for the model of a two electron
temperature plasma with a single cold ion species was derived by
\cite{DoubleBoltz} as
\begin{equation}\label{sagd}
S(\varphi,M) = M^2\left[1 - \left(1 - \frac{2\varphi}{M^2}\right)^{1/2}\right]
    + \frac{f}{\alpha_c}[1 - \exp(\alpha_c \varphi)]
    + \frac{1 - f}{\alpha_h}[1 - \exp(\alpha_h \varphi)].
\end{equation}
The derivation is straightforward: start from \eqref{cont}--\eqref{pois}
written in a frame co-moving with the solitary structure, and integrate the
resulting equations to obtain the energy integral
\begin{equation}\label{ener}
\frac{1}{2} \left( \od{\varphi}{\chi} \right)^2 + S(\varphi,M) = 0.
\end{equation}
The new parameter is the Mach number $M=V/c_{ia}$, where $V$ is the soliton
velocity.
The co-moving coordinate introduced here,
\begin{equation}\label{chi}
\chi = x - Mt,
\end{equation}
[\cite{Buti,BharShu,Baboolal,DoubleBoltz}] is similar to $\zeta$, but not
limited to slightly supersonic solitons.

It is clear from \eqref{sagd} that $S(\varphi,M)$ is limited for positive
$\varphi$ by $M^2/2$, whereas in principle there are no constraints on the
negative side.
The limitation at $\varphi=M^2/2$ comes from an infinite compression of the
cold ion density, and if one wants to obtain a soliton solution, a positive
root of $S(\varphi,M)$ must be encountered before $M^2/2$ is reached.
From $S(M^2/2,M)=0$ a maximum value $M=M_c$ is obtained, although at $M_c$ the
root is not an acceptable solution since the ion density would be infinite.

Now, insert the critical values \eqref{numval} and rewrite $S(\varphi,M)$ as
\begin{eqnarray}\label{sagdC}
S(\varphi,M) &=& \frac{5-2\sqrt{6}}{6}\, \{1 - \exp[(3+\sqrt{6})\varphi]\}
    + \frac{5+2\sqrt{6}}{6}\, \{1 - \exp[(3-\sqrt{6})\varphi]\} \nnn
&& +\, M^2\left[1 - \left(1 - \frac{2\varphi}{M^2}\right)^{1/2}\right].
\end{eqnarray}
As usual, charge neutrality in the undisturbed plasma far from the nonlinear
structure and suitable integration constants imply that $S(0,M)=S'(0,M)=0$,
where the prime denotes the derivative of $S(\varphi,M)$ with respect to
$\varphi$.
At the next stage, $S''(0,M)=0$ yields the acoustic Mach number.
Here, $M_s=1$, as a result of the normalization and conditions \eqref{numval}
on the supercritical composition, serving at the same time as the lowest
possible value for $M$.

The next stages lead to $S'''(0,M_s) = S^{(4)}(0,M_s)=0$, translating
effectively into $B=C=0$ from the KdV analyses, and $S^{(5)}(0,M_s)=24$,
showing that only positive polarity (i.e., compressive) solitons are possible.
The terminology compressive or rarefactive depends on how one chooses to define
this notion for plasmas with more than two constituents, as it is then no
longer unambiguous.

The conclusion about the soliton polarity is an extension of the result that in
generic plasmas the sign of $S'''(0,M_s)$ determines the sign of $\varphi$,
i.e., the polarity of the KdV-like solitons [\cite{TwoIonsCairns}].
By ``KdV-like" we mean that their amplitudes vanish at the true acoustic speed
and increase monotonically with the increment in soliton speed over the
acoustic speed, but these solitons might reach appreciable amplitudes, not
limited by the KdV constraints imposed by the reductive perturbation analysis.
Sometimes solitons of the opposite polarity can be generated for the same set
of compositional parameters, in addition to the KdV-like solitons, but these
cannot be obtained from reductive perturbation theory, only through a Sagdeev
pseudopotential treatment.

As a check on the link between the Sagdeev pseudopotential and reductive
perturbation approaches, we expand \eqref{sagdC} to fifth order in $\varphi$,
replace in the third- and higher-order terms $M$ by $M_s=1$, but in the
second-order term put $M=M_s+W=1+W$ and retain only the linear terms in $W$.
The rationale for this procedure is that the solitons are now slightly
supersonic, as they should be in KdV theory, but that higher order
contributions are already small enough so that the correction in $W$ is no
longer important.
Putting it all together, we obtain from \eqref{ener} that
\begin{equation}\label{expanSagd}
\frac{1}{2} \left( \od{\varphi}{\zeta} \right)^2 - W \varphi^2
    + \frac{1}{5}\, \varphi^5 = 0,
\end{equation}
having replaced $\chi$ by $\zeta$ and provided $\varphi$ is interpreted as
$\varphi_1$.
It is now straightforward to check that the solution to \eqref{expanSagd} is
precisely \eqref{onesol}, again setting $\varphi=\varphi_1$.
Analogous connections can be found elsewhere for other KdV related problems
[\cite{VerheestASSL}].
\begin{figure}
\centering
\includegraphics[width=130mm]{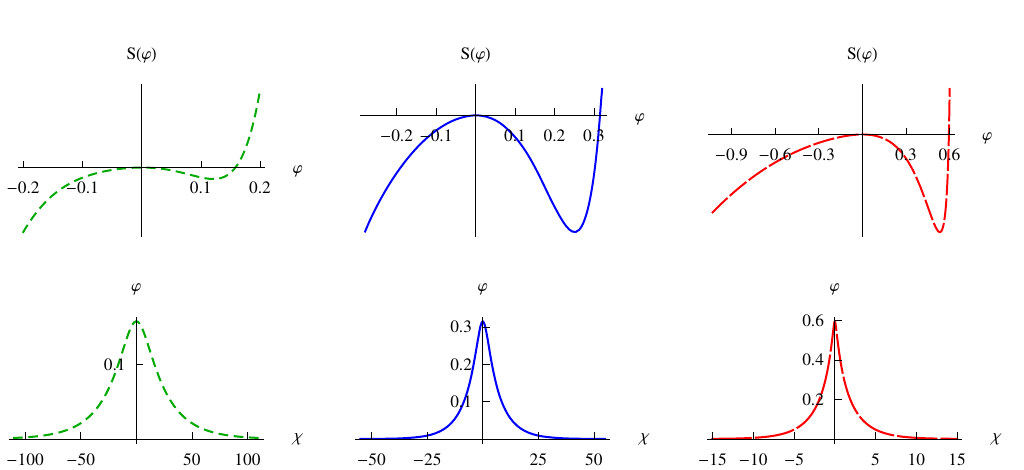}
\caption{
(Color online)
Plots of Sagdeev pseudopotentials (upper row) and their soliton solutions
(lower row), going from left to right for $M=1.001$, $M=1.01$ and $M=1.1$, with
amplitudes 0.158, 0.315 and 0.595, respectively.}
\label{SagdSol}
\end{figure}

Returning to numerical examples drawn from \eqref{ener} and \eqref{sagdC}, we
see in Fig.\ \ref{SagdSol} that the soliton amplitudes increase with $M$, until
a maximum for $M$ is reached at $M_c=1.149$, when $\varphi=0.660$, beyond which
$S(\varphi,M)$ and the cold ion density are no longer real.
At the same time, the soliton widths decrease with $M$, so that taller solitons
are narrower and faster, although one can no longer express these relations
analytically in contrast to what was possible for the supercritical KdV soliton
\eqref{onesol}.

A comparison of Figs.\ \ref{kdvSol} and \ref{SagdSol} is interesting because at
$M=1.001$ (equivalent to $W=0.001$) the KdV soliton amplitude is slightly
larger than the one obtained under the more complete Sagdeev solution.
This is also the case for $M=1.01$ or $W=0.01$.
Moreover, although not shown in Fig.\ \ref{kdvSol}, for $M=1.1$ the KdV soliton
amplitude would be 0.794, which exceeds the validity limits of the reductive
perturbation \textit{Ansatz} as well as the maximum 0.660 that the Sagdeev
formalism allows, when keeping the nonlinear terms in full without restriction.
This is, once again, a salutary reminder that KdV results have to be used and
interpreted with great care, which unfortunately is lacking in many
applications where graphs are included.

We will now explore the accuracy of the KdV solitons in more detail.
Numerical results show that the KdV equation consistently overestimate the
soliton amplitudes.
\begin{figure}
\hspace*{10mm}
\includegraphics[width=50mm]{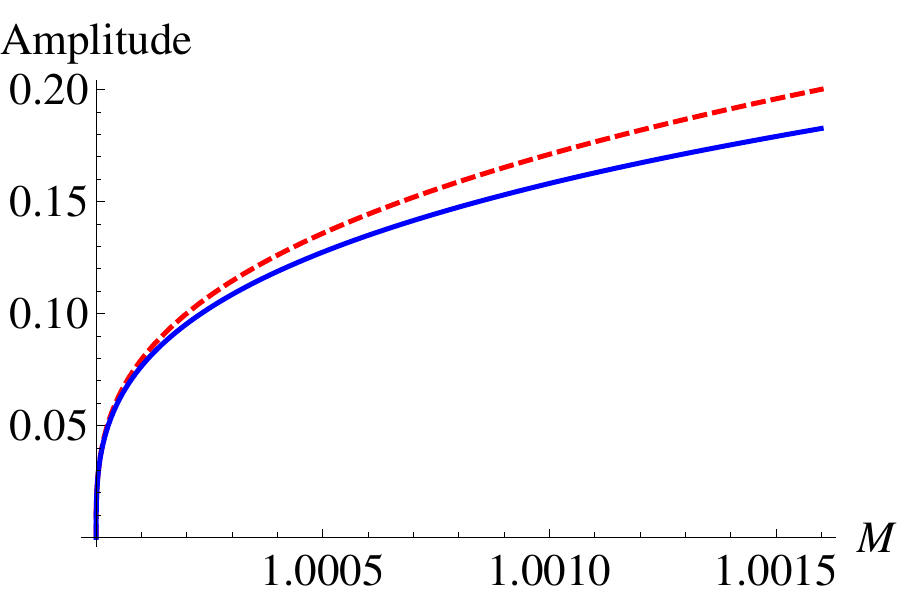} \hfill
\includegraphics[width=50mm]{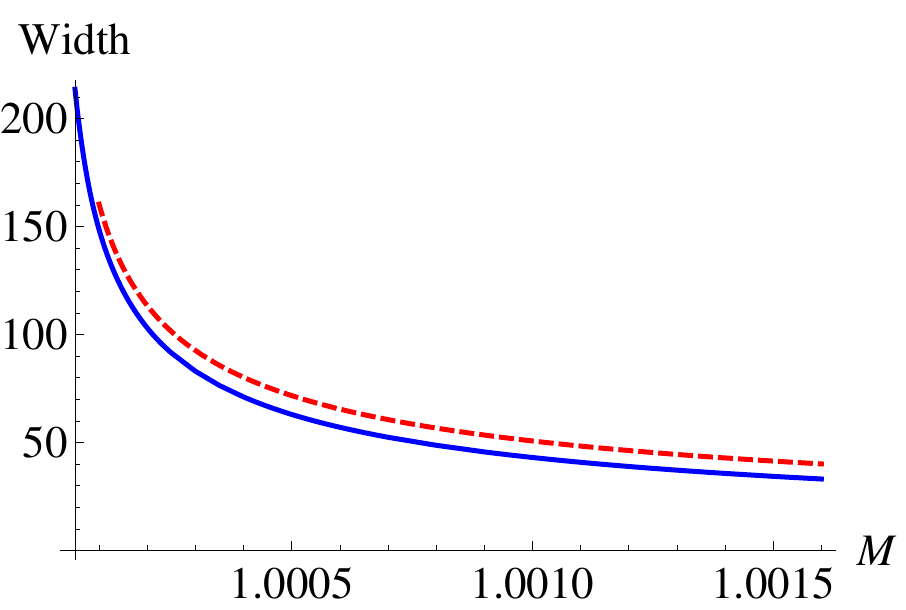}
\hspace*{10mm}
\caption{Comparison between the amplitudes (left) and widths (right) for
solitons obtained from the Sagdeev pseudopotential (solid lines) and the
modified KdV (dotted lines).}
\label{Comparisons}
\end{figure}
In Fig.\ \ref{Comparisons} (left) the amplitudes of the solitons obtained from
the Sagdeev pseudopotential (solid line) and from the KdV equation
(dotted line) are shown.
There is reasonable agreement for velocities $M<1.0002$, where the soliton
amplitude is below $0.1$.
As the velocity increases beyond $M>1.0002$, the estimate becomes more and more
inaccurate.

The accuracy of the widths of the solitons obtained from the KdV equation is
also considered.
In Fig.\ \ref{Comparisons} (right) we show the widths of the solitons obtained
from the Sagdeev potential (solid line) and from the KdV equation (dotted line).
Once again, the results agree for smaller velocities $M<1.0002$, while larger
velocities result in larger inaccuracies.

It is interesting to note that the KdV solitons consistently overestimate both
the amplitude and width of the soliton.
Also, the KdV approximation applies only to velocities that slightly exceed
the acoustic speed.

\section{Conclusions}

In this paper we have investigated the supercritical composition of a plasma
model with cold singly-charged positive ions in the presence of a
two-temperature electron population, starting initially from a reductive
perturbation approach.
The combined requirement that the evolution equation of the KdV family be free
of quadratic and cubic nonlinearities leads to a unique choice for the set of
compositional parameters and a modified KdV equation with a quartic nonlinear
term.
We believe that the model adopted here is one of the simplest that can sustain
supercriticality, but the discussion of its properties is in terms of the
structure of the modified KdV equation, rather than the precise values of its
coefficients.
Even though the present model might be difficult to generate in practice, the
conclusions will be valid for more complicated plasma compositions with some
free adjustable parameters remaining in the model equations.

Once the quartic modified KdV equation was derived, we discussed and plotted
its one-soliton solution and computed the conserved densities.
Only three of those have been found.
Consequently, the equation is not completely integrable, which precludes
finding multi-soliton solutions.
The solution is merely a solitary wave, without the elastic interaction
properties expected from solitons.

Next, since the full Sagdeev pseudopotential method had already been worked
before, with completely different focus and aims, it was straightforward to
adjust it for the chosen set of parameters and plot the corresponding fully
nonlinear solutions.
As expected, the soliton widths decrease with their velocities, so that taller
solitons are narrower and faster.
In contrast to the supercritical KdV solitons for which an analytic expression
was readily computed, one can no longer express these relations analytically,
hence, one has to rely on numerical results.

All this allows for an interesting comparison between the KdV and Sagdeev
results, which shows that the KdV solitons have slightly larger amplitudes than
those obtained under the more complete Sagdeev solution.
Only for solitons which are slightly superacoustic does the KdV analysis yield
acceptable amplitudes.
With respect to full solutions, this is, once again, a salutary reminder that
KdV results have to be used and interpreted with great caution, which is
unfortunately not always the case in many applications where graphs are
included.


\end{document}